\documentclass[twoside,12pt]{article}
\usepackage{amsmath}
\usepackage[psamsfonts]{amssymb}
\usepackage{amsthm}
\def\comm#1#2{\left[ #1,#2 \right] }

\def\Id{{\rm Id}}

\newtheorem{Thm}{Theorem}[section]

\newtheorem{Lemma}[Thm]{Lemma}
\newtheorem{theorem}[Thm]{Theorem}

\newtheorem{lemma}[Thm]{Lemma}

\def\Pref#1{Proposition~\ref{#1}}
\def\Cref#1{Corollary~\ref{#1}}
\def\Lref#1{Lemma~\ref{#1}}
\def\Tref#1{Theorem~\ref{#1}}

\def\Ad{{\rm Ad}}

\def\bH{{\mathbf H}}

\def\bbf{{\mathbf f}}
\def\bbg{{\mathbf g}}

\def\mbG{{\mathbf \Gamma}}

\def\calc{{\mathcal C}}

\def\calh{{\mathcal H}}

\def\calt{{\mathcal T}}

\def\calp{{\mathcal P}}

\def\fd{{\mathfrak D}}

\def\Z{{\mathbb Z}}
\def\R{{\mathbb R}}

\def\pslash{{\fd}}
\def\Epslash{\hbox{{\rm {\hbox{$\partial$\kern-1.2ex \raise.14ex\hbox{/}\kern.5ex}\kern-.6ex \lower.28ex \hbox{$_E$}}}}}
\def\Epslashi{\hbox{{\rm {\pslash\kern-.6ex \lower.28ex \hbox{$_{E,i}$}}}}}
\def\Epslashii#1{\hbox{{\rm {\pslash\kern-.6ex \lower.28ex \hbox{$_{E,#1}$}}}}}
\def\Epslasht{\hbox{{\rm {\pslash\kern-.6ex \lower.28ex \hbox{$_{E,\calt}$}}}}}
\def\Epslashtstar{\hbox{{\rm {\pslash\kern-.6ex \lower.28ex \hbox{$_{E,\calt^*}$}}}}}
\def\Epslashti{\hbox{{\rm {\pslash\kern-.6ex \lower.28ex \hbox{$_{E,\calt_i}$}}}}}

\def\mbni#1{\medbreak \noindent {\bf #1}\enspace}

\def\ep{\epsilon}

\def\hensp#1{\enspace\hbox{#1}\enspace}
\def\qsp#1{\qquad\hensp{#1}}

\def\la{\lambda}

\def\part{\partial}

\def\be{\begin{equation}}
\def\ee{\end{equation}}
\def\beq{\begin{eqnarray}}
\def\eeq{\end{eqnarray}}
\def\nn{\nonumber \\ }

\font\runningheadfont=cmcsc10

\def\lrp#1{\left( #1\right)}
\def\lrp#1{\left( #1\right)}
\def\lra#1{\left\langle #1\right\rangle}

\newif\ifMarginNotes \MarginNotestrue
\def\mrgn#1{\ifMarginNotes\setbox0=\vtop{\hsize 6.75pc
   {\noindent\relax #1\par}}\leavevmode
   \vadjust{\dimen0=\dp0 \dimen1=\ht0\advance\dimen1 by .5ex
 \advance\dimen0 by -.5ex
  \kern-\dimen1\hbox{\kern\hsize\kern.5pc$\leftarrow$
  \box0}\kern-\dimen0}\fi}
\catcode`\@=11 \font\twelvemsb=msbm10 scaled 1200
\font\tenmsb=msbm10 \font\ninemsb=msbm7 scaled 1200
\newfam\msbfam
\textfont\msbfam=\twelvemsb  \scriptfont\msbfam=\tenmsb
  \scriptscriptfont\msbfam=\ninemsb
\def\msb@{\hexnumber@\msbfam}
\def\Bbb{\relax\ifmmode\let\next\Bbb@\else
 \def\next{\errmessage{Use \string\Bbb\space only in math
mode}}\fi\next}
\def\Bbb@#1{{\Bbb@@{#1}}}
\def\Bbb@@#1{\fam\msbfam#1}
 
 \font\twelveeufm=eufm10 scaled 1200
\font\teneufm=eufm10 
\font\seveneufm=eufm7 
\newfam\eufmfam
\textfont\eufmfam=\twelveeufm \scriptfont\eufmfam=\teneufm
\scriptscriptfont\eufmfam=\seveneufm
\def\frak{\relax\ifmmode\let\next\frak@\else
 \def\next{\errmessage{Use \string\frak\space only in math mode}}\fi\next}
\def\frak@#1{{\frak@@{#1}}}
\def\frak@@#1{\fam\eufmfam#1}

\catcode`\@=12

\title{An Exchange Identity for Non-linear Fields}
\author{Arthur Jaffe and Christian J\"akel\footnote{Current Address: Departamento
de Fisica Matematica,
Universidade de S\~{a}o Paulo, Brasil}\\
Harvard University\\
Cambridge, MA 02138, USA}
\date{\today}

\begin{document}
\hsize=7truein \hoffset=-.75truein \maketitle \thispagestyle{empty}
\begin{abstract}
We establish a  useful identity for intertwining a creation or annihilation
operator with the heat kernel of a self-interacting bosonic field theory.
\end{abstract}
\section{Background}
Consider creation operators $a^*(f)$ and annihilation operators
$a(h)$, both linear in their respective test functions $f,h\in
L^2(\R,dx)$, acting on the Fock Hilbert space $\calh$, and
satisfying the canonical commutation relations
    \be
        \comm{a(h)}{a^*(f)}=\lra{\bar h,f}_{L^2}\;.
    \ee
The free field Hamiltonian $H_0$ acts on $\calh$ and also on the one
particle subspace $L^2(\R,dx)$, where one denotes its action by the
operator $\omega=\lrp{-d^2/dx^2+m^2}^{1/2}$. The operators $a^*, a$,
and $H_0$ satisfy the relation
    \be
        e^{ a(h)} e^{-\beta H_0} e^{a^*(f)}
        = e^{-\lra{\bar h,e^{-\beta\omega}f}}
        e^{a^*(e^{-\beta\omega}f)}
        e^{-\beta H_0}
        e^{ a(e^{-\beta\omega}  h)}\;.
    \label{FreeExchange}
    \ee
This identity (in case either $f=0$ or $h=0$) is known in the
constructive field theory literature as a ``pull-through'' identity.

The pull-through identity played a central role in the analysis of
properties of heat kernels for field theories with interaction. It
provided a fundamental ingredient in the analysis of the domain of
the fields, in the proof of the cluster expansion, in the proof of
the existence of a mass gap, and especially in the proof of the
existence of an upper mass gap in weakly-coupled $\la\calp(\varphi)$
quantum field models, see \cite{GJ4, GJS}.  An introduction to this
work can be found in \cite{QP, London}, but one must visit the
original literature for details.  The free-field pull-trough
identity provides a key step in the proof of the nuclearity property
for the {\em free} field by Buchholz and Wichmann \cite{BW}, and
motivates finding the related identity \eqref{Exchange1} for a field
theory with a $\calp(\varphi)$ polynomial interaction.

\setcounter{equation}{0}
\section{The Main Result}
In this paper we give a new identity similar to
\eqref{FreeExchange}, but with $H_0$ replaced by the Hamiltonian $H$
for a non-linear field theory (with a spatial cutoff).   Because of
the non-linearity, the Hamiltonian on the left of the identity
differs from the Hamiltonian on the right.  Remarkably, we present a
closed form for the relationship between the time-dependent
Hamiltonians.

At least one of the Hamiltonians must depend on time, so we allow
both to do so (in a particular way) and denote the time-dependent
Hamiltonians that arise by $\bH(s)$. One must replace the semigroup
$e^{-\beta H}$ by the time-ordered exponential
${T\exp{\lrp{-\int_0^\beta \bH(s) ds}}}$, where we use the
convention that time increases from left to right. Call the
resulting identity that generalizes \eqref{FreeExchange} an {\em
exchange identity}. It has the structure
    \be
        e^{ a(h)}  \lrp{T e^{-\int_0^\beta \bH_1(s)ds}} e^{a^*(f)}
        = e^{-\lra{\bar h,e^{-\beta\omega}f}}
        e^{a^*(e^{-\beta\omega}f)}
        \lrp{T e^{-\int_0^\beta \bH_2(s)ds}}
        e^{ a(e^{-\beta\omega}  h)}\;.
    \label{FullExchange}
    \ee
We give the explicit form of $\bH_1(s)$ and $\bH_2(s)$  in
\Tref{Theorem:ExchangeIdentity}.

In this paper we emphasize the algebraic structure of the exchange
identity. We do not analyze the convergence of exponential series or
the convergence of families of such series.  We expect that most
such questions in specific applications of interest can be addressed
by the reader---hopefully without undue difficulty.  In order to
ensure stability we do assume that the basic interaction polynomial
is bounded from below. In order to avoid infra-red problems we also
assume that the mass of $H_0$ is strictly positive, or else we work
with a twist field defined on a spatial circle.  All in all, the
complete justification of \Tref{Theorem:ExchangeIdentity}, even for
an elementary non-linearity, requires the introduction and removal
of an ultra-violet cutoff, using for instance, a Feynman-Kac
representation and estimates on path space to establish stability
bounds and convergence of associated vectors and operators. See the
methods in \cite{QP}.  Once one establishes the basic stability
bound in a particular example---uniform in the ultra-violet
cutoff---details concerning convergence of vectors and operators,
domains on which \Tref{Theorem:ExchangeIdentity} applies, etc., will
all fall into place.  The case of complex functions $f$ or $h$ leads
to non-hermitian Hamiltonians $\bH_1$ or $\bH_2$.  But these always
arise as small non-hermitian perturbations of a self-adjoint
Hamiltonian, so standard methods should apply.

While these steps need to be carried out in particular examples,
including such details here would obscure the simplicity of the
presentation of our new identity. This elegant form of the exchange
identity raises the question whether one might make progress toward
finding other useful closed-form expressions in the solution of
$\calp(\varphi)_2$ quantum field theories.

\subsection{Interactions}
Define a vector space of sequences of complex-valued, bounded
functions on $\R$. These vectors $\bbf\in\calc$ have components
$\bbf_j(x)$, $j\in\Z_+$. There is a natural scalar multiplication by
smooth functions $\la(x)$,
    \be
        \lrp{\la\bbf}_j(x)
        = \la(x)\bbf_j(x)\;.
    \label{ScalarMult}
    \ee
 There is also a natural imbedding $\iota:L^2(\R)\mapsto\calc$
given by
    \be
        \iota(f) = \{0,f,0,\ldots\}\;.
    \ee
In addition to multiplication \eqref{ScalarMult} by scalars, the
vector space $\calc$ is a commutative ring with the product
$*:\calc\times\calc\to\calc$ defined by
    \be
        \lrp{\bbf * \bbg}_{j}(x)
        = \sum_{k=0}^j \bbf_k(x)\bbg_{j-k}(x)\;.
    \ee
The identity in $\calc$ is the function
    \be
        \Id
        = \{1,0\ldots\}\;,
    \ee
and the $n^{\rm th}$ $*$-power of $\iota(f)$ is
    \be
        \iota(f)*\iota(f)*\cdots*\iota(f)
        = \{0,0,\ldots,f(x)^n,\ldots\}\;.
    \ee
Also define $\iota(f)^{0}=\Id$. In terms of these powers, there is a
natural exponential imbedding $\Gamma: f\mapsto\calc$ given by
    \be
        {\mbG}(f)
        = e^{\iota(f)}
        = \Id+\sum_{j=1}^\infty \frac{1}{j!}\; \iota(f)^j
        = \{1,f(x),\frac{1}{2!} f(x)^2, \ldots,
        \frac{1}{j!}f(x)^j,\ldots\}\;.
    \ee
With this notation,
    \be
        \mbG(f)*\mbG(g)=\mbG(f+g)\;,
        \qquad
        \mbG(f)^{-1}=\mbG(-f)\;,
        \qsp{and}
        \mbG(0)=\Id\;.
    \label{GammaComposition}
    \ee

The usual interaction polynomial arises from a polynomial
$\calp(\xi)$ and is defined as
    \be
        H_I(\calp,\la)
        = \int {:}\calp(\varphi(x)){:}\la(x)dx\;,
        \qsp{where}
        \calp(\xi)
        = \xi^{2k} + \sum_{j=0}^{2k-1} c_j \xi^j\;,
    \label{HInt}
    \ee
where ${:}\calp(\varphi(x)){:}$ is the normal-ordered energy
density, see for example \cite{QP}. We take the spatially dependent
cutoff $0\le \la(x)$ to be smooth and compactly supported. This
cutoff defines an interesting class of polynomial interactions.  The
polynomial $\calp$, paired with a vector $\la\bbg\in\calc$ define an
associated polynomial interaction by the relation
    \be
        \bH_I(\calp,\la\bbg)
        := \sum_{j=0}^\infty
        H_I(\calp^{(j)},\la\bbg_j)\;,
    \label{Interaction}
    \ee
where $\calp^{(j)}$ denotes the $j^{\rm th}$ derivative of $\calp$
and $\bbg_j$ the $j^{\rm th}$ component of $\bbg$.   The sum in
\eqref{Interaction} terminates with $j=2k$, the degree of $\calp$.
The special case $H_I(\calp,\la)$ of \eqref{HInt} corresponds to
$\bbg=\Id=\mbG(0)$. We use a bold-face Hamiltonian to denote one
determined by a polynomial $\calp$ (bounded from below) as well as
its derivatives $\calp^{(j)}$ in the fashion \eqref{Interaction}
with
    \be
        \bbg
        =\mbG(g)\;.
    \label{InteractionG}
    \ee
In the following we find that perturbations of this type play a
special role, especially when $g$ has the form $f_s+h_{\beta-s}$,
where
    \be
        f_s(x) = \lrp{\lrp{2\omega}^{-1/2}e^{-s\omega}f}(x) \;.
    \ee
(Note $f_0\neq f$.) Therefore consider the time-dependent, total
Hamiltonians at time $s$ of the form
    \be
        \bH
        = \bH(\calp,\la\mbG(g_s+h_{\beta-s}))
        = H_0 + \bH_I(\calp,\la\mbG(g_s+h_{\beta-s}))\;,
    \ee
(where the vacuum energy has not been renormalized to zero).

An elementary pull-through identity has the form
    \be
       e^{a(h)} \lrp{T e^{-\int_0^\beta \bH(\calp,\la\mbG(g_s))ds}}
       = \lrp{T e^{-\int_0^\beta \bH(\calp, \la\mbG(g_s+h_{\beta-s}))\,ds}}
       e^{a(e^{-\beta\omega}h)}\;.
    \label{SpecialCase}
    \ee
We establish this and related identities in the next section.

\subsection{Exchange Identities}
The following generalization states how to exchange the position of the product
of an exponential of a creation and an exponential of an annihilation operator.

    \begin{theorem} \label{Theorem:ExchangeIdentity} {\bf (Exchange Identity)}
    As a formal identity,
    \be
        e^{ a(h)}  \lrp{T e^{-\int_0^\beta \bH(\calp,\la\mbG(g_s))ds}} e^{a^*(f)}
        = e^{-\lra{\bar h,e^{-\beta\omega}f}}
        e^{a^*(e^{-\beta\omega}f)}
        \lrp{Te^{-\int_0^\beta \bH(\calp, \la\mbG(f_s+g_s+h_{\beta-s})ds}}
        e^{ a(e^{-\beta\omega}  h)}\;.
    \label{Exchange}
    \ee
    \end{theorem}

\mbni{Remark.}  The exchange identity \eqref{Exchange} reduces to
the pull-through identity \eqref{SpecialCase} for $f=0$.
Furthermore, the special choice $\bbg=\Id$ gives
    \be
        e^{ a(h)}  {e^{-\beta \lrp{H_0+ H_I(\calp,\la)}}} e^{a^*(f)}
        = e^{-\lra{\bar h,e^{-\beta\omega}f}}
        e^{a^*(e^{-\beta\omega}f)}
        \lrp{Te^{-\int_0^\beta \bH(\calp, \la\mbG(f_s + h_{\beta-s})ds}}
        e^{ a(e^{-\beta\omega}  h)}\;.
    \label{Exchange1}
    \ee
This special case shows that if one begins with a time-independent
interaction, the exchange identity gives rise to  a time-dependent
Hamiltonian.  After the exchange, the perturbation of the original
Hamiltonian involves perturbations of lower degree than $\calp$, and
the coupling constant of the highest degree term is unchanged.
Therefore the standard stability bounds of constructive quantum
field theory (based on the Feynman-Kac formula) should yield the
existence of the time ordered exponential $\lrp{Te^{-\int_0^\beta
\bH(\calp, \la\mbG(f_s + h_{\beta-s})ds}}$ of the time-dependent
Hamiltonian.

\begin{lemma} \label{Lemma:TimeOrderedExp} Let $t_1\le t_2$.  Consider the Hamiltonian
$\bH(\calp,\la\mbG(g_s))$ and the time-ordered exponential
    \be
        R(t_2,t_1) = Te^{-\int_{t_1}^{t_2} \bH(\calp,\la\mbG(g_s))ds}\;,
    \label{Rts}
    \ee
with time increasing from left to right.  Then $R(t_2,t_1)$ is the solution to
the differential equation
    \be
        \frac{\partial}{\partial t_2} R(t_2,t_1)
        = - \bH(\calp,\la\mbG(g_{t_2})) R(t_2,t_1)\;,
        \qsp{with}
        R(t,t) = I\;,
    \label{RtsEquation}
    \ee
as well as the equation
    \be
        \frac{\partial}{\partial t_1} R(t_2,t_1)
        =  R(t_2,t_1)\bH(\calp,\la\mbG(g_{t_1}))\;,
        \qsp{with}
        R(t,t) = I\;.
    \label{LtsEquation}
    \ee
\end{lemma}

\mbni{Proof.}  Assume that the time-ordered exponential \eqref{Rts}
can be expanded according to usual perturbation series. Integrating
the relation \eqref{RtsEquation} gives
    \beq
        R(t_2,t_1) &=& I - \int_{t_1}^{t_2}ds_1 \bH(s_1)R(s_1,t_1)\nn
        &=&  I - \int_{t_1}^{t_2}ds_1 \bH(s_1) + \int_{t_1}^{t_2}ds_1 \int_{t_1}^{s_1}ds_2 \,\bH(s_1)\bH(s_2)R(s_2,t_1)\nn
        &=& \cdots = \sum_{j=0}^\infty (-1)^j \int_{t_1\le s_j \cdots \le s_2 \le  s_1\le t_2} ds_1\cdots ds_j \,\bH(s_1)\cdots\bH(s_j)\nn
        &=& T e^{-\int_{t_1}^{t_2}\bH(s)ds}\;.
    \eeq
This also shows that $R(t_2+\ep,t_1)-R(t_2,t_1)\sim -\ep H(t_2)R(t_2,t_1)$. One
completes the proof that the time-ordered exponential satisfies the equation
\eqref{RtsEquation} by removing the regularization and establishing convergence
of the approximation.

A similar iteration gives
    \beq
        R(t_2,t_1) &=& I - \int_{t_1}^{t_2}ds_1 R(t_2,s_1)\bH(s_1)\nn
        &=&  I - \int_{t_1}^{t_2}ds_1 \bH(s_1) + \int_{t_1}^{t_2}ds_1
            \int_{s_1}^{t_2}ds_2 \,R(t_2,s_2)\bH(s_2)\bH(s_1)\nn
        &=& \cdots = \sum_{j=0}^\infty (-1)^j \int_{t_1\le s_1 \le s_2 \cdots \le
             s_j\le t_2} ds_1\cdots ds_j \,\bH(s_n)\cdots\bH(s_1)\nn
        &=& T e^{-\int_{t_1}^{t_2}\bH(s)ds}\;,
    \eeq
leading to \eqref{LtsEquation}.

\begin{Lemma} \label{Lemma:HICommutator}
The interaction $\bH_I(\calp,\la\mbG(g_{s}))$ satisfies
    \be
        {\bH_I(\calp,\la\mbG(g_{s}))}{e^{a^*(e^{-t\omega}f)}}
        = e^{a^*(e^{-t\omega}f)}\bH_I(\calp,\la\mbG(f_t+g_{s}))\;.
    \label{YfsIdentity}
    \ee
The corresponding relation for an annihilation operator is
    \be
       {e^{a(e^{-t\omega}h)}} {\bH_I(\calp,\la\mbG(g_{s}))}
        = \bH_I(\calp,\la\mbG(h_t+g_{s})) \,  e^{a(e^{-t\omega}h)}\;.
    \label{AnnihilationPullThrough}
    \ee

\end{Lemma}

\mbni{Proof.}  Denote $\bH_I(\calp,\la\mbG(g_{s}))$ by $\bH_I(s)$.  Then
    \be
        \comm{\bH_I(s)}{e^{a^*(e^{-t\omega}f)}}
        = e^{a^*(e^{-t\omega}f)} \lrp{e^{-a^*(e^{-t\omega}f)}
        \bH_I(s)e^{a^*(e^{-t\omega}f)} - \bH_I(s)}\;.
    \label{CommutatorExpansion}
    \ee
But
    \be
        \comm{-a^*(e^{-t\omega}f)}{\bH_I(s)}
        = -\Ad_{a^*(e^{-t\omega}f)}\lrp{\bH_I(s)}
        = H_I(\calp,\la\mbG(g_s)*\iota\lrp{f_t})\;.
    \ee
Expanding the exponential $e^{-a^*}\bH_I(s)e^{a^*}$ in
\eqref{CommutatorExpansion} as a series in $\lrp{-\Ad_{a^*}}^j$, one obtains
    \beq
        \comm{\bH_I(s)}{e^{a^*(e^{-t\omega}f)}}
        &=&  e^{a^*(e^{-t\omega}f)}\sum_{j=1}^N \frac{1}{j!}
            \lrp{-\Ad_{a^*(e^{-t\omega}f)}}^j\lrp{\bH_I(s)}\nn
        &=& e^{a^*(e^{-t\omega}f)}\sum_{j=1}^\infty \frac{1}{j!}
            H_I(\calp,\la\mbG(g_s)\lrp{*\iota\lrp{f_t}}^j)\nn
        &=& e^{a^*(e^{-t\omega}f)}
            \lrp{H_I(\calp,\la\mbG(g_s)*\mbG(f_t))-H_I(\calp,\la\mbG(g_s))}\nn
        &=& e^{a^*(e^{-t\omega}f)} \lrp{H_I(\calp,\la\mbG(g_s+f_t))
            -H_I(\calp,\la\mbG(g_s))}\;,
    \eeq
where we use \eqref{GammaComposition}.  Thus we obtain \eqref{YfsIdentity} as
claimed.  A similar argument establishes the corresponding relation
\eqref{AnnihilationPullThrough}.

\mbni{Proof of \Tref{Theorem:ExchangeIdentity}.}  Let us begin by establishing
the case $h=0$, namely
    \be
        \lrp{T e^{-\int_0^\beta \bH(\calp,\la\mbG(g_s))ds}} e^{a^*(f)}
        = e^{a^*(e^{-\beta\omega}f)}
        \lrp{Te^{-\int_0^\beta \bH(\calp, \la\mbG(f_s+g_s))ds}}\;.
    \label{ExchangeOther}
    \ee
Consider the function
    \be
        G(s') =R(\beta,s')e^{a^*(e^{-s'\omega}f)} S(s',0)\;,
    \ee
where
    \be
        R(\beta, s')=\lrp{Te^{-\int_{s'}^\beta \bH(\calp,\la\mbG(g_{s}))ds}} \;,
        \qsp{and}
        S(s',0)=\lrp{Te^{-\int_0^{s'} \bH(\calp,\la\mbG(g_{s}+f_{s}))ds}}\;.
    \ee
The left and right sides of \eqref{Exchange} equal respectively $G(0)$ and
$G(\beta)$. We compute the derivative of $G(s)$ and show that it vanishes,
proving \eqref{Exchange}.  In fact using \Pref{Lemma:TimeOrderedExp}, along
with the relation
    \beq
        \frac{d}{ds} e^{a^*(e^{-s\omega}f)}
        &=& -a^*(\omega e^{-s\omega}f)e^{a^*(e^{-s\omega}f)}
        = - \comm{H_0}{a^*(e^{-s\omega}f)} e^{a^*(e^{-s\omega}f)}\nn
        &=& - \comm{H_0}{e^{a^*(e^{-s\omega}f)}}\;,
    \eeq
we find that
    \beq
        &&\hskip-.45in \frac{d}{ds}G(s')\nn
        &=&  R(\beta,s')\lrp{\bH(\calp,\la\mbG(g_{s'})) e^{a^*(e^{-s'\omega}f)}
            -  \lrp{\frac{d}{ds'} e^{a^*(e^{-s'\omega}f)} }
            - e^{a^*(e^{-s'\omega}f)}\bH(\calp,\la\mbG(g_{s'}+f_{s'}))} S(s',0)\nn
        &=&    R(\beta,s') \lrp{{\bH_I(\calp,\la\mbG(g_{s'}))}{e^{a^*(e^{-s'\omega}f)}}
                  - e^{a^*(e^{-s'\omega}f)} \bH_I(\calp,\la\mbG(g_{s'}+f_{s'}))}
                  S(s',0)\;.
     \eeq
Using  \Lref{Lemma:HICommutator}, we infer that $dG(s)/ds=0$ as claimed and
\eqref{ExchangeOther} holds.

Next consider the case $f=0$, which we analyze by taking the adjoint of the
case established above, but with $\bar g$ in place of $g$ and $\bar h$ in place
of $f$.  This gives
    \be
        e^{ a(h)}  \lrp{A e^{-\int_0^\beta \bH(\calp,\la\mbG(g_s))ds}}
        = \lrp{Ae^{-\int_0^\beta \bH(\calp, \la\mbG(h_s+g_s))ds}}
        e^{ a(e^{-\beta\omega}  h)}\;,
    \label{ExchangeAdjoint}
    \ee
where $A$ denotes anti-time ordering.  Replacing $s$ by $\beta-s$ in the
integrands is equivalent to the replacement of anti-time-ordering  by
time-ordering.  Therefore,
    \be
        e^{ a(h)}  \lrp{T e^{-\int_0^\beta \bH(\calp,\la\mbG(g_{\beta-s}))ds}}
        = \lrp{Te^{-\int_0^\beta \bH(\calp, \la\mbG(h_{\beta-s}+g_{\beta-s}))ds}}
        e^{ a(e^{-\beta\omega}  h)}\;.
    \label{ExchangeAdjoint2}
    \ee
In order to combine the two expressions, replace $g_{\beta-s}$ by  $g_s$ to
yield,
    \be
        e^{ a(h)}  \lrp{T e^{-\int_0^\beta \bH(\calp,\la\mbG(g_{s}))ds}}
        = \lrp{Te^{-\int_0^\beta \bH(\calp, \la\mbG(h_{\beta-s}+g_{s}))ds}}
        e^{ a(e^{-\beta\omega}  h)}\;.
    \label{ExchangeAdjoint3}
    \ee
Multiply this identity on the right by $e^{a^*(f)}$.  Then move this
exponential to the left in the right-hand term: use the canonical commutation
relations to commute $e^{a^*(f)}$ past $e^{a(e^{-\beta\omega}h)}$.  Then apply
the exchange identity \eqref{ExchangeOther} that was already proved. This
yields \eqref{Exchange} and completes the proof of the theorem.

\end{document}